\documentclass[a4paper,fleqn,usenatbib]{mnras}

\usepackage[T1]{fontenc}
\usepackage{ae,aecompl}
\usepackage{graphicx}   
\usepackage{amsmath}    
\usepackage{amssymb}    
\usepackage{xcolor}    

\newcommand{\rbsc}{\hbox{1RXS\,J142001.8$-$493554}}
\newcommand{\source}{\hbox{PKS\,B1416$-$493}}

\newcommand{\fos}{\hbox{PKS\,B2152$-$699}}

\newcommand{\chandra}{\textit{Chandra}}
\newcommand{\rosat}{\textit{ROSAT}}
\newcommand{\rxte}{\textit{RXTE}}

\newcommand{\xmm}{\textit{XMM-Newton}}

\newcommand{\atca}{\textit{ATCA}}
\newcommand{\frI}{\hbox{FR\,I}}
\newcommand{\fr}{\hbox{FR\,I/II}}
\newcommand{\frII}{\hbox{FR\,II}}

\title[4.6-keV cluster around \source]
{A non cool-core 4.6-keV cluster around the bright
 nearby radio galaxy \source}

\author[D.M. Worrall \& M. Birkinshaw]
 {D.M. Worrall
and M. Birkinshaw
\\
HH Wills Physics Laboratory, University of Bristol, Tyndall
Avenue, Bristol BS8~1TL \\
}

\begin{document}

\label{firstpage}

\maketitle

\begin{abstract}

We present new X-ray (\chandra) and radio (\atca) observations of the
$z=0.09$ radio galaxy \source, a member of the southern equivalent of
the 3CRR sample.  We find the source to be embedded in a previously
unrecognized bright $kT=4.6$-keV non cool-core cluster.  The discovery
of new clusters of such high temperature and luminosity within $z=0.1$
is rare.  The radio source was chosen for observation based on its
intermediate \fr\ morphology.  We identify a cavity coincident with
the northeast lobe, and excess counts associated with the southwest
lobe that we interpret as inverse Compton X-ray emission.  The jet
power, at $5.3 \times 10^{44}$ erg s$^{-1}$, when weighted by radio
source density, supports suggestions that radio sources of
intermediate morphology and radio power may dominate radio-galaxy
heating in the local Universe.

\end{abstract}


\section{Introduction}
\label{sec:intro}

The last two decades have seen increasing awareness that radio jets
are efficient conveyors of momentum flux and AGN energy to large
distances: simulations show the need for such an extra heat source on
cosmological scales \citep[see review by][]{somerville-dave}, and
\chandra\ and \xmm\ have provided spectacular demonstrations of radio
galaxy/gas interactions in individual sources \citep[reviewed
  by][]{mcnamara, fabian}.

Observations have found many X-ray gas cavities bored by radio lobes,
and often the energy expended, scaled by AGN duty cycle, appears
sufficient to replenish that lost to radiative cooling
\citep*[e.g.,][]{birzan, dunn, panagoulia}.  How the energy is
dispersed, if favorable heating and cooling feedback cycles are to be
established, is more uncertain.  The most widely discussed ideas tend
to involve the buoyant rise of gas cavities, subsonic inflation, and
sound waves \citep[see review by][]{mcnamara-nulsen}.

Radio sources at redshift $z < 0.1$ can be studied with high linear
resolution and are the best-suited for investigating the interactions
between the relativistic jet plasma and the external (X-ray-emitting)
medium.  Of these, low-power sources of the \frI\ class are the most
numerous and widely studied.  Detection of X-ray synchrotron emission
from their jets has become common with \chandra\ \citep{worrall01}
\citep[and see review by][]{wrev}.  The mechanical power of the radio
source can be estimated through its work on the X-ray gas by combining
the enthalpy of lobe cavities with cavity age \citep{birzan}.  The
values found are used as a proxy for total jet power, $P_{\rm
  jet}$. While $P_{\rm jet}$ dominates the power in radiation-mediated
channels, it has nevertheless been found to correlate with radiative
power, $P_{\rm rad}$ \citep{birzan08}, with the latest work suggesting
slopes at the steeper extreme of earlier values: $P_{\rm jet}$ roughly
proportional to $P_{\rm rad}^{0.7}$ \citep{cavagnolo, osullivan}.
\citet{godfrey} have argued that since the correlations plot power on
both axes, common distance spreading, forcing correlations to slopes
approaching unity, may be a strong factor driving the results.
However, the main sample on which the correlation is founded uses
radio observations to whatever depth is needed to match observed X-ray
cavities, and so should be unbiased (P.~Nulsen 2016, private
communication).  In any case, as new radio-source cavities are
discovered they can be tested against the claimed
jet-power/radiative-power correlation, and, pushing to sources of
somewhat higher power, the steeper correlation (albeit with its broad
dispersion) seems to be holding up \citep[e.g.,][]{croston444, wfos}.
Earlier wisdom had been that strong shocks are not important in the
dynamics and heat deposition of radio sources \citep{mcnamara}, and it
is noteworthy that this situation has been seen to change at somewhat
higher radio powers, where the work done in driving shocks dominates
cavity power alone, for example in both 3C\,444 and
\fos\ \citep{croston444, wfos}.

When radio power is weighted by source density the correlation between
jet power and radio power implies that radio sources lying close to
the break in the radio luminosity function \citep[e.g.,][]{best}
should provide the greatest contribution of jet power in the local
Universe.  Indeed, we estimate that half of the total jet-power output
should be from sources with radio powers within a factor of three of
$3 \times 10^{25}$ W Hz$^{-1}$ at 1.4 GHz (i.e., $L_{\rm 1.4~GHz}$
between 1.4 and 12.6 $\times 10^{41}$ erg s$^{-1}$).  Such sources
typically show morphologies intermediate between \frI\ and \frII, and
for our present purposes we define sources within this power range to
be \fr\ boundary sources.  It is of particular interest to study the
environments and jet-gas interactions of these sources because of
their inferred disproportionately large contribution to feedback.

The 3CRR catalogue contains the brightest radio sources in the
northern sky, and within it there are 16 \fr\ boundary sources at $z <
0.1$, all observed with \chandra\ and some also with \xmm.  The
complete subsample (Birkinshaw et al, in preparation) contains two
broad-line radio galaxies where the X-ray emission is dominated by the
core, and several cases where the hot-gas environment appears to have
expanded away.  Notable low-redshift 3CRR \fr\ boundary sources where
jet-gas interactions are reported are 3C\,35 \citep*{mannering}, 3C
192 \citep{hk}, 3C\,285 \citep{hard285}, 3C\,305 \citep{hard305},
3C\,310 \citep{kraft310}, and 3C\,465 \citep*{hard465}.  Of these
both 3C\,305 and 3C\,310 have prominent atmospheres and exhibit strong
shocks: a shock with Mach number $\cal M \approx$ 1.7 is seen in 3-keV
gas around 3C\,310, and in the smaller-scale and cooler ($kT \approx
0.4$ keV) gas around 3C\,305, $\cal M \approx$ 2 is inferred.

To extend the number of \fr\ boundary sources for which heating
mechanisms can be studied we have turned to the southern hemisphere.
The MS4 catalog is the southern equivalent of 3CRR \citep{burgess,
  bha, bhb}.  Our joint \chandra\ and \atca\ programme concentrated
first on the brightest \fr\ boundary source in the catalogue, the
$z=0.0282$ radio galaxy \fos\ \citep{wfos}.  Here we found X-ray
cavities, lobe-inverse Compton emission, an atmosphere of $kT \approx
1$ keV, intermediate between those of 3C\,305 and 3C\,310, and strong
shocks bounding the lobes with $\cal M$ in the range 2.2 to 3.
\fos\ is of further interest in exhibiting a localized jet-gas
interaction region first reported in the form of a High Ionization
Cloud (HIC) \citep{tadhunter87} adjacent to the jet, and now known to
be associated with both a significant X-ray-emitting atmosphere and
jet bending \citep{wfos}.  A similar interaction is also now known for
3C\,277.3 \citep*{wandy}.  \fos\ has rounded radio lobes but with
hotspots that are embedded rather than at the extremities, a common
feature of \fr\ boundary objects \citep*{capetti}.  Its nuclear
optical spectrum is of intermediate ionization and exhibits both
forbidden and permitted lines with broad wings \citep{tadhunter88},
and so it would be classed as a high excitation radio galaxy (HERG).

In this paper we report the X-ray (\chandra) and radio (\atca)
properties of a second MS4-catalogue \fr\ boundary source, \source.
The 4.8-GHz radio map of \citet{burgess} already showed evidence for
embedded hotspots. \source\ has other radio similarities to \fos.  If
moved from its redshift of $z =0.0914$ \citep{jones09} to that of
\fos, its 408-MHz flux density would be 76.7 Jy, as compared with 61.6
Jy for \fos.  Its angular extent would be 128 arcsec, as compared with
80 arcsec for \fos. Its 1.4-GHz core flux density would be 529 mJy, as
compared with 770 mJy for \fos.  There is a difference: in contrast to
\fos\ the nuclear optical emission of \source\ shows only weak
H$_\alpha$ and [N\,II] \citep{simpson}, and it would be classed as a
low excitation radio galaxy (LERG).

X-ray emission associated with \source\ was first reported with
\rosat: the source appears in the Bright Source Catalog as
\rbsc\ \citep{voges}, and subsequently it is listed in the 3--20-keV
\rxte\ all-sky slew survey \citep{revnivtsev}. The source has
attracted little attention, perhaps because it is in the Southern sky
at a Galactic latitude of only 10.8 degrees and at $-43$ degrees
Galactic longitude.  It was not previously identified as a cluster of
galaxies, nor are there reports of a cluster based on optical
observations of the galaxy field.

In this paper we report the properties of the cluster, and the jet-gas
interactions and energetics.  We adopt values for the cosmological
parameters of $H_0 = 70$~km s$^{-1}$ Mpc$^{-1}$, $\Omega_{\rm {m0}} =
0.3$, and $\Omega_{\Lambda 0} = 0.7$.  Thus 1~arcsec corresponds to a
projected distance of 1.7~kpc
at \source.

Throughout the paper the power-law spectral index, $\alpha$, is
defined in the sense that flux density is proportional to
$\nu^{-\alpha}$.  X-ray spectral indices are quoted in terms of
$\alpha$ rather than the values one larger returned by
spectral-fitting codes.  Uncertainties correspond to 90 per cent
confidence for the parameter of interest, unless otherwise stated.

\section{Observations and reduction methods}
\label{sec:obs}%

\subsection{\chandra\ X-ray}
\label{sec:xobs}

Our observations of \source\ were with the \chandra\ Advanced CCD
Imaging Spectrometer (ACIS) in VFAINT full-frame data mode on 2015
June 3rd (OBSID 17058).  The source was positioned at the nominal
aimpoint of the I3 chip, and the three other ACIS-I chips and the
ACIS-S S2 chip also received data (see the \chandra\ Proposers'
Observatory Guide\footnote{ http://cxc.harvard.edu/proposer} for
details of the instrument and its modes).  We followed the software
`threads' from the \chandra\ X-ray Center (CXC)\footnote{
  http://cxc.harvard.edu/ciao} to re-calibrate the data with the
energy-dependent sub-pixel event repositioning algorithm applied.
Results presented here use {\sc ciao v4.8} and the {\sc caldb v4.7.2}
calibration database.

The observations were free from background flares and, after removal
of time intervals when the background deviated more than $3\sigma$
from the average value, the exposure time was 62.829 ks.  We adjusted
the astrometry by 0.36 arcsec, mostly in RA, to align the X-ray
nucleus with the radio core position measured to be RA$=14^{\rm h}
20^{\rm m} 03^{\rm s}\llap{.}704$, Dec$=-49^\circ 35' 42''\llap{.}06$
(see Section \ref{sec:robs}).

The {\sc ciao wavdetect} task was used to find sources in the
0.4--5-keV image, with the threshold set at 1 spurious source per
field.  Except for those detections corresponding to features
associated specifically with features of \source, the corresponding
regions were masked and corrected for in quantitative analyses.  All
spectral models include absorption along the line of sight in our
Galaxy with a column density of $N_{\rm H} = 1.49 \times 10^{21}$
cm$^{-2}$, as given by the CXC {\sc colden} task using data of
\citet{dlock90}. Spectral fitting uses {\sc
    xspec}\footnote{https://heasarc.gsfc.nasa.gov/xanadu/xspec/}
  version 12.9.0u with the abundance table of \citet{anders}.

For analysis where we sample the background using blank-sky fields, in
contrast to using a local off-source measure of the background, we
followed procedures described in the CXC software threads.  We cleaned
the source and sky-background data using the same criteria, and
re-projected the background data to the coordinate system of the
source.  A small downwards correction factor of 3.5\% was applied to
the background rates so that the on-source and background rates match
at 9.5--12-keV where the particle component of the background
dominates.

\subsection{\atca\ radio}
\label{sec:robs}

We observed \source\ with the \atca\ in its 6B array configuration
with the Compact Array Broadband Backend (CABB) correlator
\citep{wilson} on 2014 August 26th (programme C2936). Data were
taken in 2048 1-MHz wide channels centred at either 17~and 19~GHz
or 5.5~and 9~GHz. The planned observing schedule was modified to take
account of the wet and thundery weather conditions, so that less time
was spent at the higher frequencies than planned, and additional
calibration was performed. Some time was lost to periods of high wind
or local lightning. A summary of the on-source time achieved is given
in Table~\ref{tab:radio}, which also shows the restoring beams and
noise levels of the three maps made, centred at 5.5, 9, and
18~GHz.

\begin{table}
\caption{ATCA radio data in programme C2936}
\label{tab:radio}
\begin{tabular}{cccc}
\hline
(1) & (2) & (3) & (4) \\
Frequency
  & On-source time 
  & Restoring Beam
  & Stokes I noise \\
 (GHz) 
  & (h) 
  & (arcsec)$^2$ 
  & ($\mu$Jy beam$^{-1}$) \\
\hline
 5.5 & 2.3 & $1.99 \times 1.17$ & 26 \\
 9.0 & 2.3 & $1.29 \times 0.70$ & 21 \\
18.0 & 4.5 & $0.50 \times 0.50$ & 17 \\
\hline
\end{tabular}
\medskip
\end{table}

Sources B1921$-$293 and B1934$-$638 were observed for bandpass and
primary flux calibration, respectively. B1421$-$490, a small-angular
size source near \source, was observed for phase calibration and as a
pointing and focus reference. Data calibration was carried out with
the {\sc
  miriad}\footnote{http://www.atnf.csiro.au/computing/software/miriad}
software using the calibration advice for CABB data given in the 2010
September version of the Miriad User Guide, with the central 90\% of
each 2-GHz band being used for mapping.

Extensive data flagging was required before and after calibration to
remove interference (at 5.5~ or 9~GHz) and periods when the atmosphere
was highly unstable. At 5.5~and 9~GHz multi-frequency synthesis using
the entire band was successful at generating useful images, via the
normal self-calibration cycle. The 18-GHz map, created from the 17-GHz
and 19-GHz data, could not be made in this way because significant
time-dependent phase slopes were seen across the bandpass, presumably
from atmospheric delays. An approximate correction for this was done
by dividing each of these two bands into eight sub-bands in frequency,
and then self-calibrating each sub-band separately against the
multi-frequency synthesis map. Large amplitude changes from
uncorrected atmospheric opacity changes were also removed in this
procedure.

The noise levels in the final Stokes I maps are affected by residual
phase and amplitude errors, but these are negligible at the contour
levels presented here, and dynamic ranges of about $10^3$ are achieved
over much of the area of the images. Polarization images were also
made, and show significant features in bright parts of the radio
emission, though only a trace of polarization associated with the
inner part of the jet can be seen near the core.

\section{Results and Discussion}
\label{sec:results}

\subsection{The radio galaxy}
\label{sec:radio}%

\begin{figure}
\centering
\includegraphics[width=1.0\columnwidth]{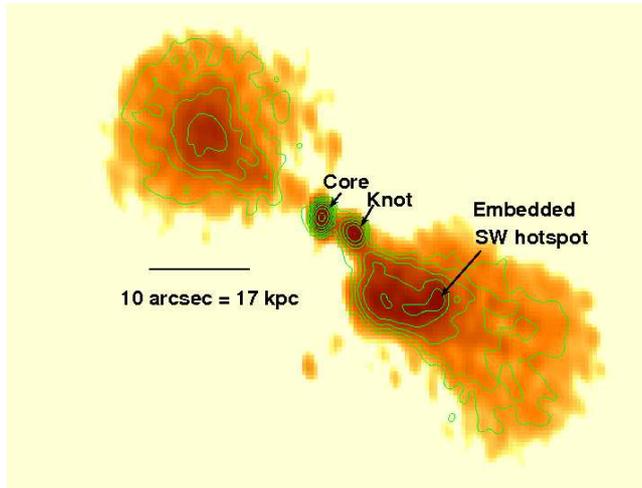}
\caption{ The 5.5 GHz radio map overlaid with contours from the 9-GHz
  map (Table~\ref{tab:radio}) showing the overall structure of the radio galaxy.
The contours increase by factors of 2 from a
lowest level of 0.2 mJy beam$^{-1}$.
}
\label{fig:radio}
\end{figure}

Our new radio mapping confirms the structure of
\source\ (Fig.~\ref{fig:radio}) to be of intermediate
\fr\ morphology. It displays two radio lobes NE and SW of the radio
core, with a jet connecting the core to the brightest part of the SW
lobe, and a faint bridge connecting the lobes. At low
surface-brightness levels both lobes appear approximately circular,
but while the bright region in the smaller NE lobe is placed
centrally, that in the SW lobe, which lies at approximately the same
distance from the core as in the NE lobe, is displaced away from the
centre of the lobe.

A prominent flat-spectrum jet knot (Table~\ref{tab:radioflux}), of
1~arcsec angular extent, lies about 3~arcsec SW of the core, and is
connected to it, and to the brightest structure in the SW radio lobe,
by fainter emission. The jet from the core to the bright knot is
misaligned with the line from the core to the brightest compact
component in the SW lobe: the radio knot appears to mark a kink in the
jet. Polarization data suggest that the magnetic field changes
direction from parallel to perpendicular to the jet axis at this
kink. Beyond the knot the jet bends again, and appears to terminate at
an interior hot spot at RA$=14^{\rm h} 20^{\rm m} 02^{\rm
  s}\llap{.}48$, Dec$=-49^\circ 35' 50''\llap{.}7$ in the SW
lobe. This hot spot lies at the end of a bright curved ridge which
extends to the E, with the apparent magnetic field running parallel to
the ridge.

\begin{table}
 \caption{Radio components}
 \label{tab:radioflux}
 \begin{tabular}{lrrrcl}
  \hline
  (1) & (2) & (3) & (4) & (5) & (6) \\
  Component & $S_{5.5 {~\rm GHz}}$
            & $S_{9{~\rm GHz}}$
            & $S_{18{~\rm GHz}}$
            & $\alpha$
            & Notes \\
            & (mJy)
            & (mJy)
            & (mJy)
            & 
            &  \\
  \hline
  Core     & $ 56 \pm 3$ & $ 56 \pm 1$ & $75 \pm 4$ & 0.00 &     \\
  Jet knot & $ 26 \pm 2$ & $ 16 \pm 1$ & $12 \pm 1$ & 0.42 & a,c \\
  NE lobe   & $221 \pm 5$ & $133 \pm 3$ & $48 \pm 3$ & 1.03 & b,d \\
  SW lobe   & $324 \pm 7$ & $196 \pm 4$ & $93 \pm 5$ & 1.02 & b,e \\
  \hline
 \end{tabular}
 \medskip
\begin{minipage}{\columnwidth}
%
%
Flux densities include systematic flux scale errors of 2\% at 5.5 and
9~GHz, and 5\% at 18~GHz because of poor weather during the
observations. \\
a. Jet knot spectrum taken from 9 to 18 GHz: 5-GHz flux
   density confused with inner lobe.\\
b. Lobe spectrum taken from 5.5 to 9 GHz: at 18 GHz the
   lobe is over-resolved, and the flux density in the table relates
   only
   to the compact inner region.\\
c. Knot angular size (Gaussian deconvolved FWHM) 1.1 x 0.2 arcsec.\\
d. NE lobe diameter about 22 arcsec, but the brightest region has a
   diameter of about 8 arcsec.\\
e. SW lobe diameter about 27 arcsec, but the brightest region has a
   diameter of about 12 arcsec.\\
\end{minipage}
\end{table}


No such well-defined structure is seen in the NE lobe, though a
fainter, possibly compact, component lies diametrically opposite the
hot spot seen in the SW lobe, at the same distance from the core. The
bright region in the NE lobe is fainter than that in the south, and is
relatively poorly represented in the present data because of the 
poor weather during the observation.

The radio core is essentially unpolarized and of flat radio
spectrum from $5.5$ to $9$~GHz, but then brightens to $18$~GHz,
suggesting the presence of a small-scale self-absorbed
component. A faint polarized extension of the core to the SW seen on
our highest-resolution 18-GHz map can be interpreted as the base of
the kpc-scale jet.

An unrelated point source lies at the edge of the SW lobe, at
RA$=14^{\rm h}
20^{\rm m} 03^{\rm s}\llap{.}84$, Dec$=-49^\circ 35' 57''\llap{.}4$.

\subsection{Large-scale X-ray emission}
\label{sec:Xextended}%

\begin{figure}
\centering
\includegraphics[width=1.0\columnwidth]{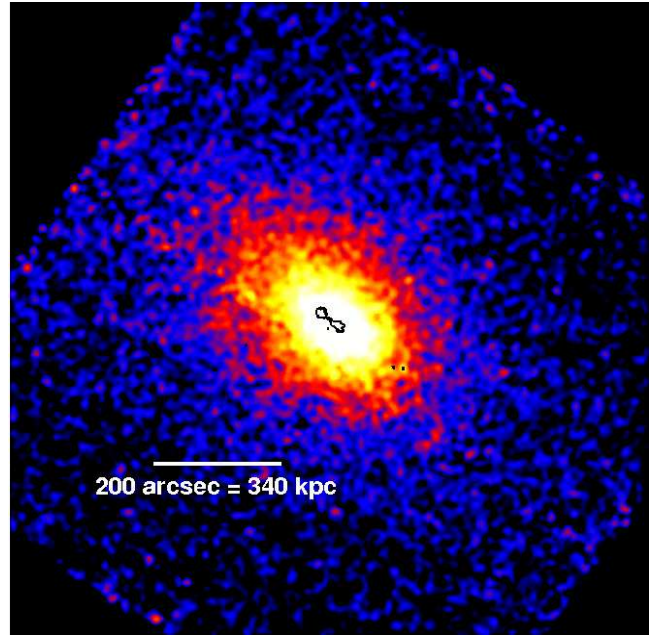}
\caption{ 0.4-5-keV exposure corrected \chandra\ image showing the lie
  of the cluster on the four ACIS-I chips.  The image is in native
  0.492-arcsec pixels and smoothed with a Gaussian of standard
  deviation $\sigma = 10$ pixels.  Point sources have been removed.
  The lowest contour from Figure~\ref{fig:radio} (black) shows that
  the radio source lies central to the cluster.}
\label{fig:acisI}
\end{figure}

From an exposure-corrected image, X-ray emission reminiscent of
cluster emission in excess of background is visually apparent over a
large fraction of ACIS-I (Fig.~\ref{fig:acisI}).  The $R = 15.7$~mag
host galaxy of \source\ \citep{jones09} can be identified as the
central brightest cluster galaxy (BCG), and we see that the radio
extent is quite well aligned in projection with the major axis of the
cluster emission.

For an initial exploration of the cluster properties, the spectrum of
the brightest part of the extended X-ray emission was measured from an
elliptical region of semi-major and semi-minor axes 52.7 and 34.3
arcsec, respectively, centered on the position of the radio core.
This on-source region lies entirely on the I3 chip, and background was
sampled from a more remote part of the same chip, to the north of the
brightest cluster emission.  The spectrum
fits well a single-temperature thermal (APEC) model ($\chi^2 = 172.9$
for 152 degrees of freedom) with $kT = 4.72^{+0.36}_{-0.35}$ keV and
abundances relative to Solar of $0.54^{+0.17}_{-0.15}$ Z$_\odot$.
This indicates a relatively rich cluster, not previously catalogued.
Similar spectral results were obtained using background from a matched
region on the blank-sky fields, processed as described in
Section~\ref{sec:xobs}.

\begin{figure}
\centering
\includegraphics[height=0.34\textwidth,clip=true]{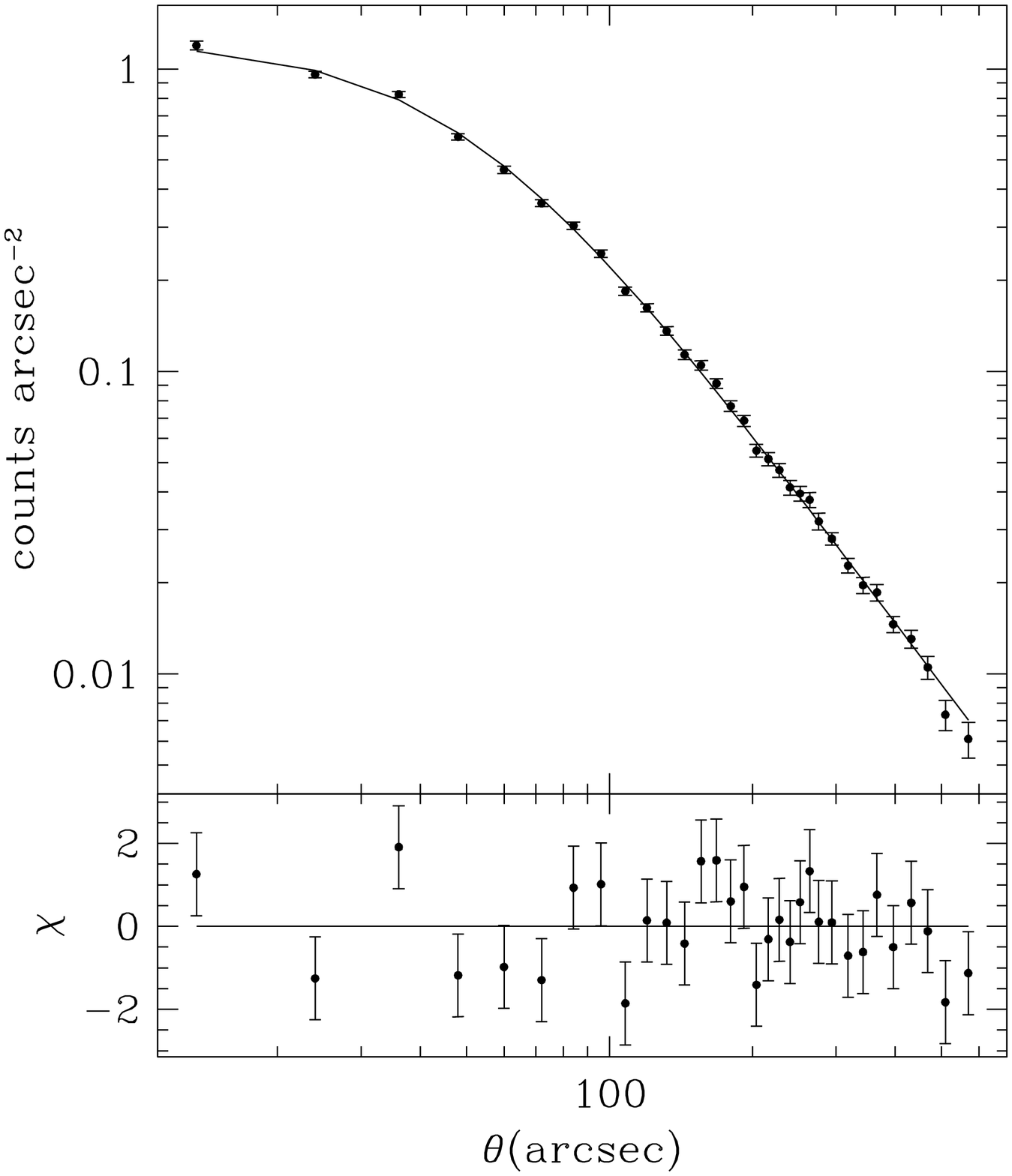}
\includegraphics[height=0.23\textwidth,clip=true]{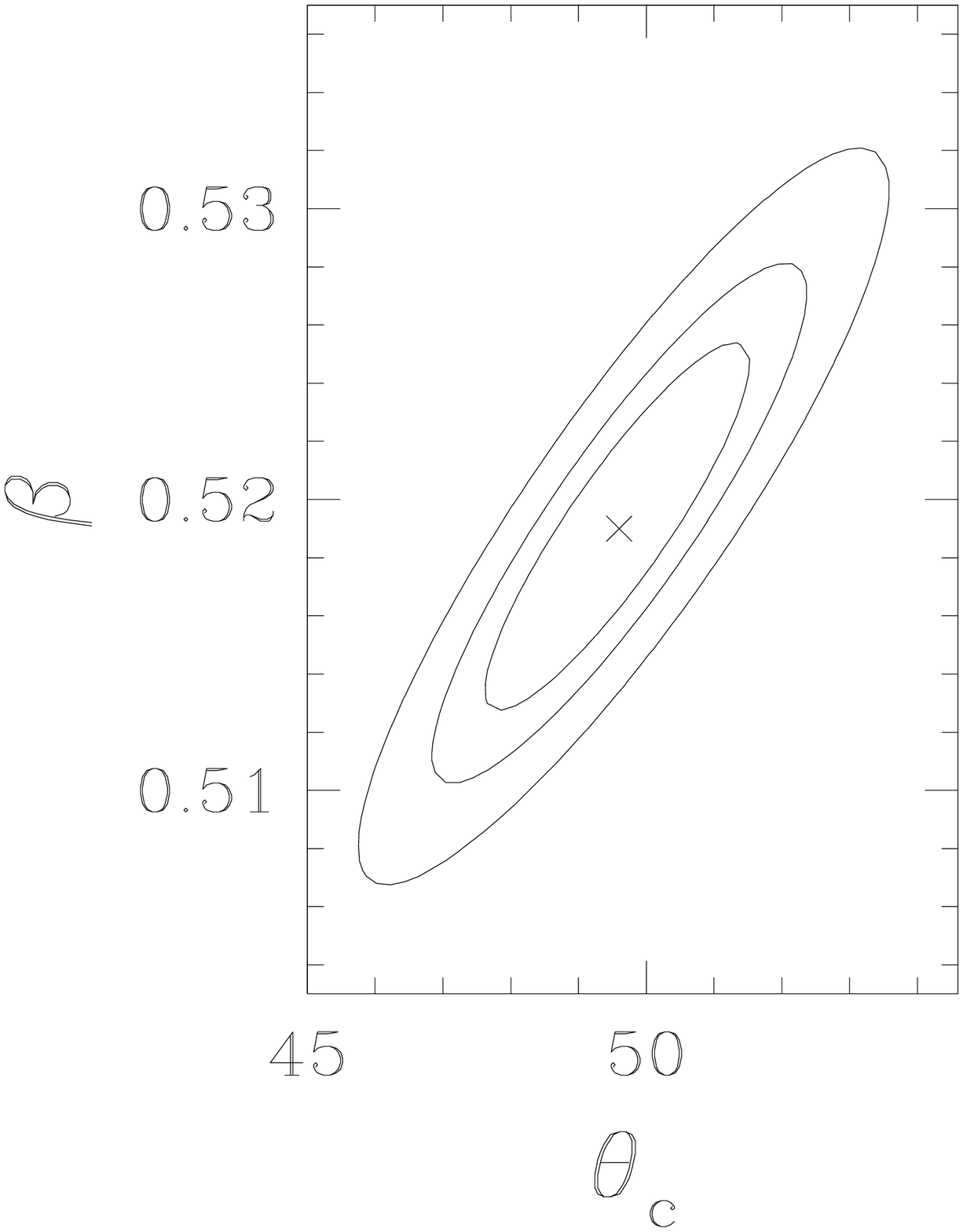}
\caption[]{{\bf Left:} Background-subtracted exposure-corrected
  0.5--7-keV radial
  profile fitted to a $\beta$-model profile with the residuals (shown
  as a contribution to $\chi$) in the lower panel.  The best fit is
  for a core radius of $\theta_{\rm c} =49.6$ arcsec and $\beta
  =0.519$ ($\chi^2/$dof = 33.7/29).  {\bf Right:} Uncertainty contours
  ($1\sigma$, 90\% and 99\% for 2 interesting parameters) of
  $\theta_{\rm c}$ and $\beta$ for the radial-profile fit.}
\label{fig:profile}
\end{figure}

\begin{figure}
\centering
\includegraphics[width=0.7\columnwidth]{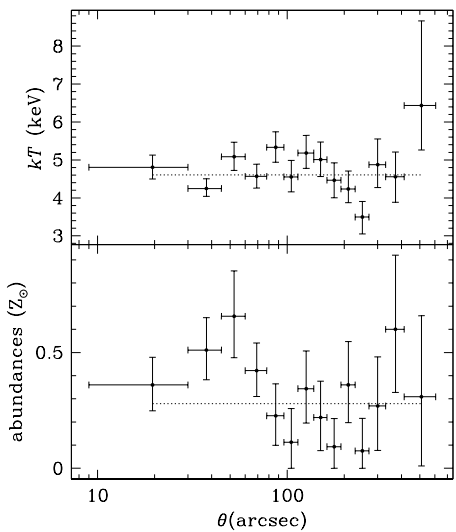}
\caption{ \chandra\ X-ray projected temperature and abundance profiles
  of the cluster showing consistency with uniformity.  Individual
  errors are $1\sigma$ for one interesting parameter.  Upper panel
  shows $kT$: average value is 4.61 keV, $\chi^2/$dof = 19.3/13.
  Lower panel shows abundances: average value is 0.28 Z$_\odot$,
  $\chi^2/$dof = 19.2/13.}
\label{fig:tprofile}
\end{figure}

Our in depth characterisation of the cluster emission uses exclusively
blank-field background subtraction.  For ease of measurement, and
because we do not know the extent of the cluster emission along the
line of sight, we have used a spherical approximation to model the
large-scale gas distribution.  Figure \ref{fig:profile} shows the
0.5--7-keV exposure-corrected radial profile, centered on the radio
core, fitted to a $\beta$ model.  The quality of fit is excellent.
Temperature and abundance profiles, constructed to provide roughly
2500 net counts in each bin, are shown in Figure~\ref{fig:tprofile}.
The quality of the individual spectral fits is acceptable at better
than 95\% confidence.  Any deviations from uniformity are small and
not highly significant, and we find average values of $kT$ and
abundance in annuli out to 10 arcmin of 4.6 keV and 0.28 Z$_\odot$,
respectively.  In particular, there is no evidence for a
cluster-scale large cool core.

For the cluster's measured temperature, the scaling relations of
\citet{vikhlinin} give a value of $r_{500}$ of 617 arcsec (1.053 Mpc),
which is also approximately the distance to which these observations
detect the cluster.  Properties of the cluster are summarized in
Table~\ref{tab:cluster}.  The values have been measured using the
radial profile (Fig.~\ref{fig:profile}) and average spectral
properties (Fig.~\ref{fig:tprofile}) using equations given in
\citet{birk} and \citet{wbrev}.

\begin{table}
\caption{Cluster Properties}
\label{tab:cluster}
\begin{tabular}{ll}
\hline
(1) & (2) \\
Parameter & Value \\
\hline
$\beta$ & $0.519\pm0.009$ \\
core radius, $r_{\rm c}$ & $84.7 \pm 4.4$ kpc \\
FWHM & $163\pm 5$ kpc \\
Average $kT$ & $4.61 \pm0.17$ keV \\
Average abundances & $0.28\pm 0.04$ Z$_\odot$\\
Bolometric $L_{r_{500}}$ & $(2.73\pm0.02) \times
10^{44}$   erg s$^{-1}$ \\
Central gas density & $(5.5\pm 0.2) \times 10^{3}$ m$^{-3}$  \\
Central pressure & $(9.3\pm 0.4) \times 10^{-12}$ Pa  \\
Central cooling time & $(1.00 \pm 0.05) \times 10^{10}$ yr \\
Mass deposition rate within $r_{\rm c}$ & $23.5\pm 1.7$ M$_\odot$ yr$^{-1}$\\
Gas mass within $r_{500}$ & $(4.05 \pm 0.25) \times 10^{13}$ M$_\odot$\\
Total mass within $r_{500}$ & $(2.79 \pm 0.12) \times 10^{14}$ M$_\odot$\\
\hline
\end{tabular}
\medskip
 \medskip
\begin{minipage}{\columnwidth}
Errors are statistical only. Unknown geometry
contributes additional uncertainties.
\end{minipage}
\end{table}

We conclude that \source\ lies in an extreme atmosphere amongst the
studied \fr\ boundary objects.  Perhaps the closest examples in 3CRR
are 3C\,388 and 3C\,338, both classified as LERGS.  3C\,388 falls just
beyond our definition of a \fr\ boundary object in the direction of
high radio power, and has been found to reside in a $kT \approx 3.5$
keV cluster with no strong cool core \citep{kraft388}.  \source\ is
more extreme in having a richer X-ray atmosphere and longer central
cooling time.  3C\,338 falls just beyond our definition of a
\fr\ boundary object in the direction of low radio power.  It lies
within the cool core of the $kT \approx 4.5$ keV cluster Abell 2199,
and both radio source and cluster are highly disturbed as a result of
merger activity \citep{nulsen}.

\citet{ineson} finds a correlation between radio power and the X-ray
luminosity of the environment for radio galaxies classed as LERGS.
The correlation is driven by a shortage of lower-power sources in rich
clusters and higher-power sources in weak atmospheres, and in a broadened
intermediate \fr\ boundary region the atmospheres span three orders of
magnitude in X-ray luminosity.  \source\ is perhaps remarkable less in
the richness of its X-ray-emitting atmosphere than in that atmosphere
having been unrecognized before now.

\subsection{Cluster ellipticity and AGN location}
\label{sec:ellipticity}%

In an attempt to characterise the ellipticity of the cluster we first
applied the {\sc iraf stsdas ellipse}
software\footnote{http://www.stsci.edu/institute/software\_hardware/stsdas}
to the smoothed data of Figure~\ref{fig:acisI}.  This provided
representations of the ellipticity and position angle of the cluster
emission on scales larger than the radio source that were used to set
starting values in subsequent fitting using the {\sc ciao sherpa}
package.  For {\sc sherpa} we binned the exposure-corrected 0.4-5~keV
image into 2-arcsec pixels, fitted over a large region of roughly 9 by
11.5 arcmin oriented with the cluster, and used the cstat statistic
because most bins contained few counts.  On top of a flat background
of fitted amplitude, an elliptical Gaussian represented the cluster
gas, and its parameters were allowed to be free.  A small-scale
elliptical Gaussian was used as a crude representation of the combined
core and galaxy emission, discussed later in more detail in Sections
\ref{sec:galaxy} and \ref{sec:core}, and a circular Gaussian of fixed
position and 1.5 arcsec FWHM was used to represent the knot emission
discussed later in Section \ref{sec:knot}. The overall fit was
acceptable, as given by the cstat criterion that the fit statistic
divided by the number of degrees of freedom should be approximately
one.

\begin{figure}
\centering
\includegraphics[width=1.0\columnwidth]{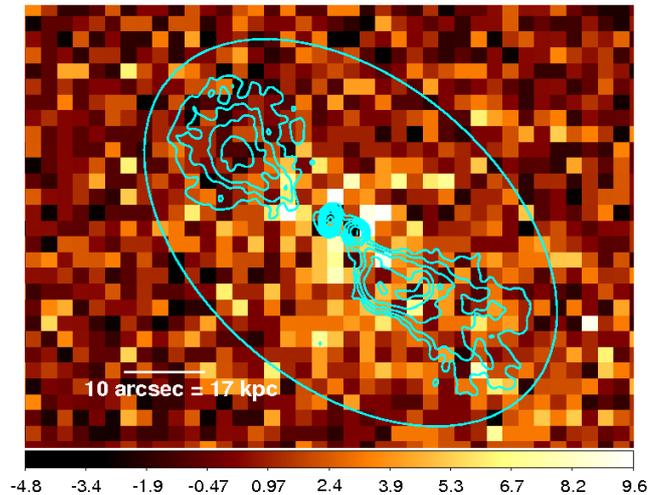}
\caption{Residual 0.4-5 keV
    exposure-corrected \chandra\ image in 2-arcsec pixels from a
    simple model
    with small elliptical and circular components at the core and
    knot, respectively, and a large elliptical component representing
    the cluster gas. Colour scale is in units of counts. 
    Overlayed are the radio contours as in
    Figure~\ref{fig:radio}, and an isophote of the cluster model to
    indicate its orientation and ellipticity.  
    The residuals show an
    asymmetry, with negative values under the NE lobe and positive
    values across the centre and in the SW.}
\label{fig:ratio}
\end{figure}

Figure~\ref{fig:ratio} shows the residual between data and model for
the inner regions. Spatially-correlated residuals show that the true
structure is more complex than the model used.  In particular, there
are correlated negatives underlying the NE radio lobe, whereas
residuals in the SW have a greater tendency to be positive.  The
central regions are generally poorly represented, with what appears to
be an excess to the NW of the radio knot that extends in a line to the
E of the SW radio lobe, and an excess bounding the inner edge of the
NW lobe.  These emissions are apparent in the original image and,
while they may represent disturbed gas, a much deeper X-ray
observation would be required to investigate morphology and origin.
We note that {\sc sherpa} fitting is restricted to conformal ellipses,
whereas visual examination and our {\sc ellipse} fitting suggest that
the cluster is more complex in spatial structure with ellipticity,
position angle, and centre varying with radius.

The elliptical component describing the cluster in the {\sc sherpa}
fit is found to be centered at RA$=14^{\rm h} 20^{\rm m} 03^{\rm
  s}\llap{.}45\pm 0.04$, Dec$=-49^\circ 35' 43''\llap{.}6\pm 0.6$.
The radio core is 2.9 arcsec (4.9 kpc) to the NE of this, and so even
without considering the third dimension, the BCG is not located
precisely at the cluster centre and relative motion is expected.  This
is also suggested by the radio structure, where the pronounced NE-SW
asymmetry is consistent with motion in a projected NE direction.  Ram
pressure would then explain the rounding of the more compact NE lobe,
whereas instabilities of flow along the side of the trailing SW lobe
would contribute to its outer plume-like appearance.

\subsection{Galaxy atmosphere}
\label{sec:galaxy}%

\begin{figure}
\centering
\includegraphics[width=0.85\columnwidth]{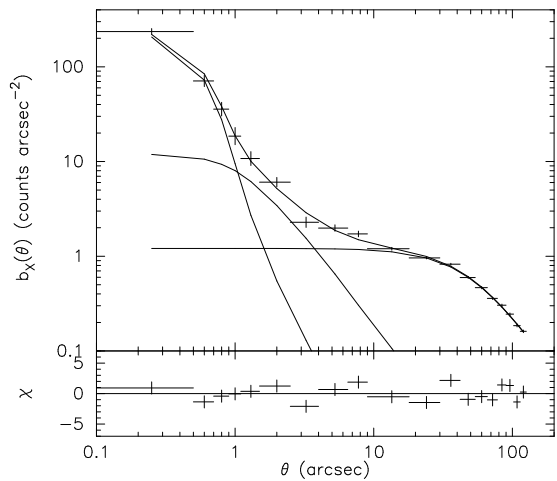}
\caption{ As Fig.~\ref{fig:profile} but reaching in to the AGN.  The
broadest model component has been fixed to the best-fit cluster values. 
The other two model components are the PSF,
representing the AGN core, and a $\beta$-model convolved with the PSF,
representing galaxy-scale gas.  The best fit finds
$\beta=0.49$, $\theta_{\rm c} = 0.97$
arcsec ($\chi^2/$dof = 27.9/16).}
\label{fig:gasradial}
\end{figure}

To investigate the X-ray emission from the host galaxy of \source\ we
extended the radial profile of Figure~\ref{fig:profile} into the core,
masking a region around the knot and truncating the cluster emission
at $126''$.  In fitting the radial profile we fixed the values of
$\beta$ and core radius representing cluster gas to values found in
Section~\ref{sec:Xextended}.  To represent the core we simulated and
merged 50 representations of the point spread function (PSF) using the
CXC {\sc SAOsac raytrace} and {\sc marx} software for the spectral
model described in Section~\ref{sec:core}, and included this with free
amplitude.  We modelled the galaxy emission as a second $\beta$-model,
but since it is of small scale it was convolved with the PSF.  The
best overall model fit, shown in Figure~\ref{fig:gasradial}, is
formally marginally acceptable, and we take it as including a
reasonable representation of galaxy-related emission in the central
region.  A deeper X-ray observation would be required to warrant more
complex modelling, which could then allow for small offsets between
cluster centre and AGN core, as discussed in
Section~\ref{sec:ellipticity}.

Guided by Figure~\ref{fig:gasradial}, spectral fitting of the
galaxy-scale emission was performed by extracting counts from a
core-centred annulus of radii 1.3 and 5 arcsec, with background from a
core-centred annulus of radii 9 and 12 arcsec.  There were too few
counts to attempt models more complex than a simple single-temperature
thermal model, and whereas cluster emission should be accounted for by
the choice of background region, contamination from the core PSF and
emission from X-ray binaries in the host galaxy of \source\ is not. We
found an adequate fit to $kT = 1.6^{+1.8}_{-0.8}$ keV ($\chi^2/$dof =
6.7/8). The best-fit abundance is 0.3 Z$_\odot$, but with
unconstrained errors. We have used the temperature together with the
fitted spatial parameters of the radial profile to derive properties
of the galaxy gas given in Table~\ref{tab:galaxy}.


\begin{table}
\caption{Galaxy gas properties}
\label{tab:galaxy}
\begin{tabular}{ll}
\hline
(1) & (2) \\
Parameter & Value \\
\hline
$\beta$ & $0.5\pm0.1$ \\
core radius, $r_{\rm c}$ & $1.7^{+1.7}_{-1.2}$ kpc \\
FWHM & $3.2^{+3}_{-2}$  kpc \\
Average $kT$ & $1.6^{+1.8}_{-0.8}$ keV \\
$L_{0.5-2~{\rm keV}} (<5'')$ & $(4.5\pm1.0) \times
10^{41}$   erg s$^{-1}$ \\
Gas density at $5''$ & $(15\pm2) \times 10^{3}$
m$^{-3}$  \\
Cooling time at $5''$ & $(2^{+1}_{-0.7}) \times
10^{9}$ yr \\
Mass deposition rate within $r_{\rm c}$ & $0.4^{+0.4}_{-0.3}$ M$_\odot$ yr$^{-1}$\\
Gas mass within $5''$   & $(1.0^{+2.5}_{-0.9}) \times 10^{10}$ M$_\odot$\\
\hline
\end{tabular}
\medskip
 \medskip
\begin{minipage}{\columnwidth}
Errors are statistical only.
\end{minipage}
\end{table}


The galaxy atmosphere is not unusual for the host of a radio galaxy
--- even with \rosat\ the techniques we have used here could separate
thermal atmospheres of galaxy scale from central AGN in the closest
radio galaxies \citep{wbpspc}, and work has been extended to many more
sources using the much narrower PSF of \chandra.  \citet{sun} refers
to galaxy-scale atmospheres as coronae, distinguishing them from
cluster cool cores of the type absent around \source, further
suggesting that the radio outbursts made possible by the presence of
coronae in turn help destroy embryonic cluster cool cores.  Radial
profiles of cool-core clusters \citep[e.g.,][]{henning} lack the flattening
and inflection between cluster and galaxy core radii seen in
\source\ (Fig.~\ref{fig:gasradial}).

\subsection{Lobes and jet power}
\label{sec:features}%

\begin{figure}
\centering
\includegraphics[width=1.0\columnwidth]{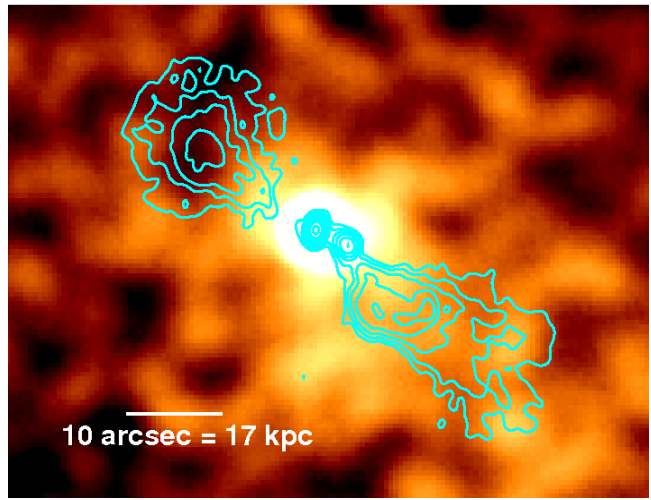}
\caption{ Contours of the 9-GHz radio map, as in
  Figure~\ref{fig:radio}, superimposed on a 0.4-5
  keV exposure-corrected \chandra\ image in native 0.492-arcsec
  pixels, where data have been smoothed with a Gaussian of $\sigma =
  4$ pixels.  There is a deficit of X-ray emission projected onto the
  NE lobe as compared with the SW.  }
\label{fig:radio-on-cluster-smo8}
\end{figure}

Figure \ref{fig:radio-on-cluster-smo8} presents radio contours on a
smoothed image of the X-ray emission.  The more rounded NE lobe
appears to lie against a deficit of X-ray emission as compared with
the SW lobe.  We have quantified this by sampling background for each
lobe from regions in sectors $90^0$ away and extending over similar
cluster-modelled elliptical annuli.  The NE lobe shows no excess
relative to the cluster at a 90\%-confidence upper limit of 0.9~nJy at
1 keV (adopting a power-law spectrum of index consistent with the
radio).  In the case of the SW lobe the excess (about 15\% of the
surrounding cluster surface brightness) gives a good fit to a
power-law spectrum of $\alpha_{\rm x} = 1.2^{+1.9}_{-1.1}$
($\chi^2/$dof= 6.7/9) with a 1-keV flux density of $4.8_{-3.3}^{+4.9}$
nJy.  The fit to a thermal model is equally good with $kT =
3.3^{+\infty}_{-2.2}$ ($\chi^2/$dof= 6.5/9).  .

\begin{figure}
\centering
\includegraphics[width=0.7\columnwidth]{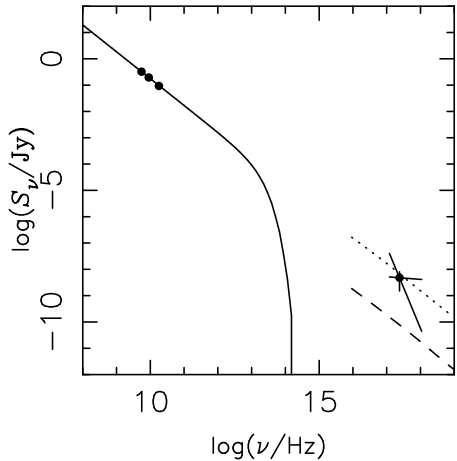}
\caption{ Radio and X-ray measurements of the SW lobe and a model of
  synchrotron emission (solid line) and inverse Compton scattering of
  the same power-law electron population on the cosmic microwave
  background (dotted line) and synchrotron self Compton emission
  (dashed line).  The model is for a magnetic field strength of 0.5
  nT, a factor of four below equipartition, which brings the
  synchrotron and external cluster pressures into agreement (Table
  \ref{tab:press}).}
\label{fig:swlobe}
\end{figure}

It is attractive to interpret the X-ray excess in the SW lobe as due
to inverse Compton scattering by the radio synchrotron-emitting
electron population, as supported by the agreement of radio and X-ray
spectral indices.  Pressure arguments back this up.  Firstly, we have
used the radio data in Table~\ref{tab:radio} to calculate the
total pressures for an equipartition magnetic field in each
lobe, following \citet*{hardsynch}, and have compared with gas
pressures at similar projected distances from the core measured
through our X-ray radial-profile and spectral fitting
(Table~\ref{tab:press}).  We find lobe underpressure by a
factor of eight (columns 3 and 4, Table~\ref{tab:press}), which can be
restored to balance by increasing or decreasing lobe field strength to
the values shown in column 5, or by adding protons.  For the SW lobe,
decreasing the field strength to a factor of four below equipartition
gives both an excellent match to the gas pressure and an
inverse-Compton prediction that matches the X-ray excess
(Fig.~\ref{fig:swlobe}).  It is of course likely that the lobe is
overpressured and the external gas shocked, as can be achieved by
adding a likely component of relativistic protons. Shocked gas is very
difficult to see in hot clusters like this one, although such gas is
measured at a Mach number of roughly 3 in the 1-keV atmosphere around
\fos\ \citep{wfos}.  There is no evidence for \source\ being at small
angle, $\theta$, to the line of sight, but if, for example,
$\theta=30^\circ$, the inverse-Compton required field of 0.5~nT would
leave the SW lobe overpressured by only a factor of 1.3, although this
is before consideration of a proton contribution.

\begin{table}
\caption{Lobe synchrotron and gas pressures}
\label{tab:press}
\begin{tabular}{lcccc}
\hline
(1) & (2) & (3) & (4) & (5)\\
Feature
  & $B_{\rm eq}$
  & $P_{\rm eq}$
  & $P_{\rm clust}$
  & $B_{\rm P_{clust}}$  \\
  & (nT) 
  & ($10^{-12}$ Pa)
  & ($10^{-12}$ Pa) 
  & (nT)  \\
\hline
NE lobe 
  & 2.1 
  & 1.2
  & $8.8\pm 0.4$
  & 0.6, 7.9   \\
SW lobe 
  & 2.0
  & 1.0
  & $8.4\pm 0.4$
  & 0.5, 7.8 \\
\hline
\end{tabular}
\medskip
 \medskip
\begin{minipage}{\columnwidth}
 Columns (2) and (3) show equipartition magnetic field strength and
 pressure, respectively.  Column (4) shows the cluster gas pressure at
 the angular distance from the core to the centre of the lobe.  Column
 (5) gives magnetic field strengths that would result in enough
 synchrotron pressure to match the cluster value.  Note that a 0.6 nT
 field in the NE lobe would violate constraints on inverse Compton
 X-ray emission unless there is a gas cavity at the lobe.  A
 0.5 nT field in the SW lobe could explain the excess X-ray seen here
 as inverse Compton emission (Fig.~\ref{fig:swlobe}).  Synchrotron
 calculations use a minimum electron Lorentz factor of $\gamma_{\rm
   min}=100$, and ignore non-uniform filling factor and protons.
\end{minipage}
\end{table}

It seems reasonable that departures from equipartition would not be
too dissimilar between opposing radio lobes, and departures in the
sense of magnetic fields being lower by factors of around three are
rather common for radio-galaxy lobes in general \citep{crostonlobes}.
The NE lobe might then be expected also to show a field below
equipartition and a small fractional excess of X-ray emission relative
to the ambient cluster, but here a field of 0.6 nT (Table
\ref{tab:press}) produces X-ray emission that exceeds our measured
90\%-confidence upper limit for any X-ray excess.  The measured
deficit in counts as compared with cluster emission is about 13\% and
can be seen by eye in Figure~\ref{fig:radio-on-cluster-smo8}.  This
suggests that, as in many other local radio sources
\citep[e.g.,][]{birzan, birzan08}, the NE lobe has evacuated its
internal X-ray emitting gas.  If a lobe lying within the core radius
of the cluster is evacuated of gas, then for our measured value of
$\beta=0.52$ we expect the surface brightness to be reduced by a
factor that is roughly the ratio of the lobe radius ($11''$) to the
core radius ($50''$), or in this case around 20\%.  Even allowing for
plausible levels of lobe inverse Compton emission, within the large
uncertainties we conclude that a lobe cavity is responsible for the
X-ray counts deficit.

This begs the question as to why no cavity is seen against the SW
lobe.  While thermal structure on scale sizes comparable to the lobes
may be influencing our results (a much longer X-ray exposure would be
required to map such structure), in Section~\ref{sec:ellipticity} we
pointed out that the NE lobe is more compact and rounded than that in
SW and there is a small projected offset from the cluster centre, both
suggesting motion of the radio galaxy within the potential well of the
cluster towards the direction of the NE lobe.  Instabilities of flow
along the side of the trailing SW lobe may then have led to greater
re-mixing of external gas in the SW.

We have employed the usual method of estimating the cavity power,
$P_{\rm cav}$ of the NE lobe as $4 P_{\rm clust} V/t_{\rm s}$ where
$P_{\rm clust}$ is the external pressure (Table~\ref{tab:press}), $V$
is the volume (we adopt the full $22''$ lobe diameter from
Table~\ref{tab:radioflux}), and $t_{\rm s}$ is the lobe crossing time
at the external sound speed \citep{birzan}.  Doubling the value, to
account for both jets, gives a proxy for the jet power averaged over
the buoyancy time of 32 Myr as $P_{\rm jet} = 5.3 \times 10^{44}$ erg
s$^{-1}$.  The total 1.4-GHz radio flux density of \source\ at 1.4 GHz
is 2.6 Jy \citep{wright}, giving a radio power of $P_{\rm 1.4~GHz} =
7.3 \times 10^{41}$ erg s$^{-1}$.  These measurements are within the
scatter of the correlation between $P_{\rm jet}$ and $P_{\rm 1.4~GHz}$
reported by \citep{birzan, birzan08} and updated by \citet{cavagnolo}
for radio sources in cooling-flow clusters.  \source\ does not require
kinetic and thermal energy of shocked gas to bring it into agreement,
as is necessary for the bright well-studied southern \fr\ radio galaxy
\fos\ \citep{wfos}. $P_{\rm jet}$ is larger than the bolometric
luminosity of the cluster gas out to a radius of $r_{500}$
(Table~\ref{tab:cluster}), and so the jet is easily able to provide
enough energy to reverse a cooling flow on the scale of the cluster
core radius.

The source \source\ then adds to the known number of cavity radio
sources, and is one of a smaller group where lobe inverse Compton
emission is also detected, so that lobe energetics are measured.  The
magnetic field is below the equipartition value, and while shocks are
not detected, their presence, along with a relativistic or cold
entrained proton population, cannot be excluded.  The fact that like
\fos, \source\ lies on a $P_{\rm jet} \propto P_{\rm rad}^{0.7}$
correlation supports the hypothesis that \fr\ boundary sources, when
weighted by radio-source density, should dominate radio-galaxy heating
in the local Universe.  The fact that these radio sources are observed
in atmospheres with a range of richness, with that of \source\ being
among the richest, means that heating is not confined to one
particular type of environment.

\begin{figure}
\centering
\includegraphics[width=1.0\columnwidth]{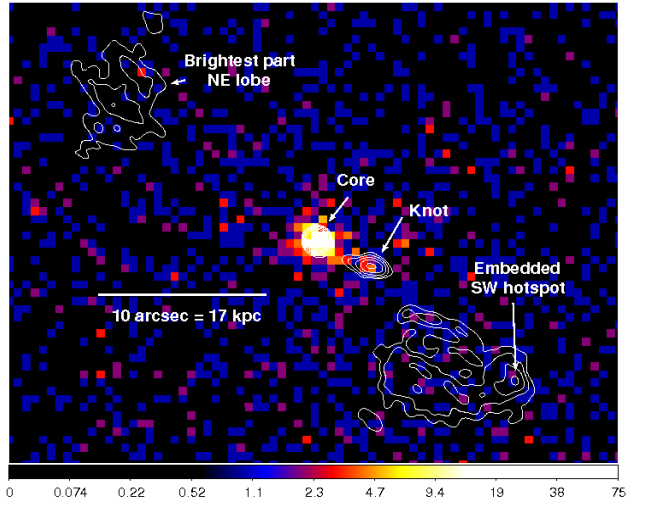}
\caption{ Contours of the 18-GHz radio map (Table \ref{tab:radio})
  superimposed on unsmoothed 0.4-5 keV X-ray data in native
  0.492-arcsec pixels. Colour scale is in units of counts. The
  contours increase by factors of 2 from the lowest level of 0.2 mJy
  beam$^{-1}$.  Excess X-rays are seen at the core and radio knot but
  are not apparent in the brightest part of the NE lobe or the
  embedded SW hotspot.}
\label{fig:radio-on-cluster}
\end{figure}

\subsection{Core}
\label{sec:core}%

\begin{figure}
\centering
\includegraphics[width=0.7\columnwidth]{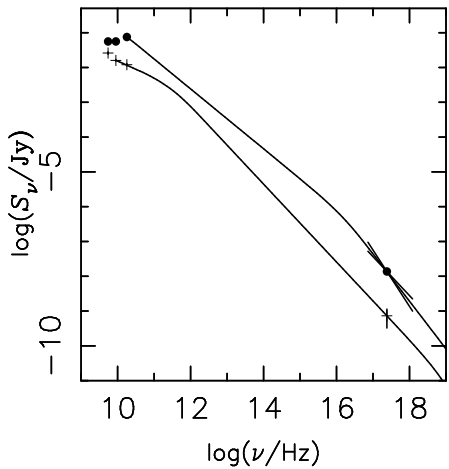}
\caption{ Radio and X-ray measurements of the core (circles) and jet
  knot (crosses) with possible broken-power law synchrotron spectral
  models. For the core $\Delta\alpha=0.5$ is possible; inverse-Compton
  emission cannot be modelled because the region is unresolved and the
  size unknown.  For the jet $\Delta\alpha>0.5$ is needed
  ($\Delta\alpha=0.7$ shown); implausibly the field would need to be
  almost 100 times below equipartition to give instead the jet X-rays via
  inverse-Compton (synchrotron self-Compton) scattering of the radio
  synchrotron-emitting electron population.  }
\label{fig:core-knot}
\end{figure}

The core of \source, whose X-ray counts are seen in Figure
\ref{fig:radio-on-cluster}, has a radio spectrum that is rising at 18
GHz, indicative of self-absorption.  The X-ray spectrum (containing
roughly 300 counts) was measured from a circle of radius 1.25 arcsec,
and was found to be well modelled with a relatively steep power law of
$\alpha_x = 1.36^{+0.26}_{-0.24}$ and no excess absorption.  Pileup is
less than 1\%, which justifies ignoring the contribution of
out-of-time events from the core to the cluster emission.  The 0.2-10
keV luminosity is $2.7 \times 10^{42}$ erg s$^{-1}$, and the flux
density at 1 keV is $13.7 \pm 2.4$ nJy.  The spectral index
interpolated between the radio (18 GHz) and X-ray is $\alpha_{\rm rx}
= 0.95$, typical of jet cores and knots \citep*[e.g.,][]{balmaverde,
  harwood}.  If the X-ray is interpreted as dominated by emission from
a small-scale jet and arising from the electron population responsible
for the radio core, the combination of $\alpha_x$ and $\alpha_{\rm
  rx}$ requires the synchrotron spectrum to steepen somewhere between
the radio and X-ray, and a spectral break of $\Delta\alpha = 0.5$ is
the simple prediction for energy losses (although observations of
resolved jet knots often find somewhat larger breaks
\citep*[e.g.,][]{bm87, hard66b, birkins}.  The data for \source\ can
be accommodated by a break at the relatively high frequency of about
$10^{16}$ Hz if $\Delta\alpha=0.5$ between a presumed index of
$\alpha_{\rm r} = 0.86$ in the radio and $\alpha_{\rm x} = 1.36$
(Fig.~\ref{fig:core-knot}). The break frequency would be lower for a
larger value of $\Delta\alpha$.

The X-ray emission of \source's core is thus characteristic of LERG nuclei, with
the X-rays inferred to arise from the base of a jet.  No excess
absorption is required.

\subsection{Jet knot}
\label{sec:knot}%

The radio properties of the jet knot marked in Figures~\ref{fig:radio}
and \ref{fig:radio-on-cluster} are given in Table~\ref{tab:radioflux}.
The X-rays have been measured from a knot-centred circle of radius 1.5
arcsec, with the cluster and other background sampled from a
core-centred annulus of radii 2.06 and 5.06 arcsec, excluding the knot
region.  There are roughly 17 net counts.  Statistical errors dominate
the uncertainty in 1-keV flux density; adopting $\alpha=1.1$
($\Delta\alpha=0.7$ larger than the radio spectral index, see Tab.~
\ref{tab:radioflux} and below), we
find a value of $0.75^{+0.45}_{-0.36}$ nJy.

The lower points in Figure~\ref{fig:core-knot} show the radio and
X-ray flux densities of the knot.  Given the relatively small source
size, predictions for X-ray inverse-Compton radiation are dominated by
the synchrotron self-Compton process, but the field strength would
need to be almost 100 times below the equipartition value of 14~nT to
match observations, with an implausible increase in total energy.
Instead it is more likely that the X-rays are synchrotron, as inferred
to be common in jet knots of \frI\ radio galaxies \citep{worrall01}, 
found not unusual in \frII\ radio galaxies \citep[e.g.,][]{wb346, kraft403,
  kataoka, hardpica}, and favoured for some quasar jets
\citep[e.g.,][]{jester07, cara, meyer}.  The precise characteristics of the radio to X-ray
spectrum cannot be determined in the absence of high-resolution
observations at intermediate frequencies, but we can infer that the
spectral break needs to be $\Delta\alpha > 0.5$ to accommodate the
measured value of $\alpha_{\rm r}$ and the interpolated value of
$\alpha_{\rm rx}$, and $\Delta\alpha=0.7$ is adopted in
Figure~\ref{fig:core-knot}.  We also note that the registration
between X-ray and radio knots is not good, with the X-rays
peaking a fraction of a detector pixel upstream of the radio.  This
behaviour has been seen in other radio-galaxy knots for which a
synchrotron origin is inferred
 \citep[e.g.,][]{hard66b, wb346}.

Figure \ref{fig:radio-on-cluster} shows no excess X-ray
emission associated with the embedded SW hotspot.


\section{Summary and Conclusions}


This study of \source\ was undertaken to obtain further
information on the range of gas/radio plasma interactions associated
with intermediate-power radio galaxies, since such sources should be 
responsible for a major fraction of the radio-source heating in the
Universe. \source\ is one of the brightest such \fr\ sources,
and adds to the small number of sources of this class that has been
mapped with sub-kpc linear resolution in the X-ray band.

An unexpected finding of this work is that \source\ lies within a few
kpc of the centre of a hot, dense, cluster atmosphere. The host
cluster has not previously been reported in optical or X-ray
surveys. Radio sources of distorted shape, such as 3C\,83.1B in the
Perseus cluster \citep{odeaowen}, are signposts to the presence of a
dense gas environment, but \source\ was known only as a relatively
symmetrical double source with embedded hot spots, so only a low-mass
group atmosphere was expected. \source\ joins the roughly $30\%$ of
sources at the \fr\ boundary that are known to lie in significant
groups or clusters.

%
%

The source \source\ shows several features commonly found in deep
X-ray images of intermediate-power radio galaxies. The jet displays an
X-ray- and radio-bright synchrotron knot about 7~kpc from the
core. The jet changes direction at this knot, and there is a
significant (few hundred pc) offset between the X-ray and radio
emission peaks.  The NE lobe occupies a cavity in the surrounding
X-ray gas, while we interpret excess counts in the SW lobe as
inverse-Compton X-ray emission which, with the implied modest
departure from equipartition, can bring the lobe into pressure balance
with the external medium.  The X-ray exposure is insufficiently deep
for us to address thermal structure on the scale size of the radio
source or relativistic proton content of the lobes. Only the SW lobe
shows a compact hotspot, which is (as usual when only one hotspot is
seen) located on the jet side of the source. The hotspot is recessed
from the edge of the lobe --- a feature that is sometimes interpreted
as an indication of source orientation in \frII\ sources, but may
instead be an indication of an internal disruption of the jet without
significant generation of X-rays in \fr\ sources, as it is in the
wide-angle tailed sources such as 3C\,465 \citep{hard465}.

We find no evidence for shocks around the lobes of \source.  Shocks in
atmospheres as hot as that around \source\ are difficult to detect
using current X-ray telescopes, and for a given input energy would be
expected to exhibit a lower Mach number in a hot atmosphere than in a
cool atmosphere. The clear shocks found around the similar source
\fos\ \citep{wfos} would only be seen as weak temperature and X-ray
surface-brightness changes if \fos\ were to be embedded in the
\source\ cluster.

The heat input from \source\ to the cluster
environment has been found to be consistent
with the correlation with radio source power seen in other sources
\citep[e.g.,][]{cavagnolo}, and is a factor 30 higher than
the power required to balance the cooling of gas within the cluster's 
core radius. Such strong central heating should be driving strong
convective motions in the atmosphere --- the high X-ray count rate and
large angular size of the cluster suggests that these motions could be
detectable with the next generation of X-ray telescopes.

\section*{Acknowledgments}

We thank the STFC for travel support to the \atca, and are grateful to
the helpful staff at Narrabri.  The scientific results reported in
this article are based on observations made with the Chandra X-ray
Observatory and the Australia Telescope Compact Array.  The Australia
Telescope is funded by the Commonwealth of Australia for operation as
a National Facility managed by CSIRO. We are grateful to the
\chandra\ X-ray Center (CXC) for its support of \chandra\ and the {\sc
  ciao} software.  This research has made use of the NASA/IPAC
Extragalactic Database (NED) which is operated by the Jet Propulsion
Laboratory, California Institute of Technology, under contract with
the National Aeronautics and Space Administration. We thank the
referee for constructive comments which have helped improve the
manuscript, and Leith Godfrey and Paul Nulsen for discussions of
jet-power correlations.


\end{document}